\documentclass[preprint,10pt]{aastex}

\newcommand\sqd{{deg$^{2}$}}
\newcommand\etal{{\it et al. }}
\newcommand\kms{km~s$^{-1}$}
\newcommand\msun{$M_\odot$}

\def\ncat{578}
\def\ndet{435}
\def\nprior{129}
\def\nhvc{14}
\def\be{\begin{equation}}
\def\ee{\end{equation}}

\shorttitle{Third ALFALFA Catalog}
\shortauthors{Kent et al.}

\begin{document}
\title{The Arecibo Legacy Fast ALFA Survey: \\
       VI. Second HI Source Catalog of the Virgo Cluster Region
}

\author {Brian R. Kent\altaffilmark{1},
Riccardo Giovanelli\altaffilmark{1,2}, Martha P. Haynes\altaffilmark{1,2}, 
Ann M. Martin\altaffilmark{1},
Am\'elie Saintonge\altaffilmark{3},
Sabrina Stierwalt\altaffilmark{1},
Thomas J. Balonek\altaffilmark{4},
Noah Brosch\altaffilmark{5},  
Rebecca A. Koopmann\altaffilmark{6}
}

\altaffiltext{1}{Center for Radiophysics and Space Research, Space Sciences Building,
Cornell University, Ithaca, NY 14853. {\it e--mail:} bkent@astro.cornell.edu, riccardo@astro.cornell.edu,
haynes@astro.cornell.edu, amartin@astro.cornell.edu, 
amelie@astro.cornell.edu, sabrina@astro.cornell.edu}

\altaffiltext{2}{National Astronomy and Ionosphere Center, Cornell University,
Space Sciences Building,
Ithaca, NY 14853. The National Astronomy and Ionosphere Center is operated
by Cornell University under a cooperative agreement with the National Science
Foundation.}

\altaffiltext{3}{Institute for Theoretical Physics, University of Zurich, Winterhurerstrasse 190, CH-8057 Zurich, Switzerland
{\it e--mail:} amelie@physik.uzh.ch}

\altaffiltext{4}{Dept. of Physics \& Astronomy, Colgate University, Hamilton, NY 13346.
{\it e--mail:} tbalonek@mail.colgate.edu}

\altaffiltext{5}{The Wise Observatory \& The School of Physics and Astronomy, 
Raymond \& Beverly Sackler Faculty of Exact Sciences, Tel Aviv University, Israel.
{\it e--mail:} noah@wise.tau.ac.il}

\altaffiltext{6}{Dept. of Physics \& Astronomy, Union College, Schenectady, NY 12308.
{\it e--mail:} koopmanr@union.edu}

\begin{abstract}

We present the third installment of HI sources extracted from the Arecibo
Legacy Fast ALFA extragalactic survey.  This dataset
continues the work of the Virgo ALFALFA
catalog.  The catalogs and spectra published
here consist of data obtained during the 2005 and 2006 observing sessions
of the survey.  The catalog consists of \ncat~HI detections within the
range $11^h36^m <$ R.A.(J2000) $< 13^h52^m$ and 
$+08^{\circ} <$ Dec.(J2000) $< +12^{\circ}$, and $cz_{\odot} < 18000$ \kms.
The catalog entries are identified with optical counterparts where possible
through the examination of digitized optical images.  The catalog detections
can be classified into three categories: (a) detections 
of high reliability with S/N $>$ 6.5; (b) high velocity clouds
in the Milky Way or its periphery; and (c) signals of lower S/N
which coincide spatially with an optical object and known redshift.
75\% of the sources are newly published HI detections.  Of particular note is a complex
of HI clouds projected between M87 and M49 that do not coincide with any optical
counterparts.  Candidate
objects without optical counterparts are few.  
The median redshift for this sample is 6500 \kms~ and
the $cz$ distribution exhibits the local large scale structure consisting
of Virgo and the background void and the A1367-Coma supercluster regime at
$cz_{\odot} \sim$7000 \kms.  Position corrections for telescope pointing errors are applied
to the dataset by comparing ALFALFA continuum centroid with those
cataloged in the NRAO VLA Sky Survey.  The uncorrected positional accuracy averages
 27\arcsec ~(21\arcsec ~median) for all sources with S/N $>$ 6.5 and is of order 
$\sim$21\arcsec ~(16\arcsec ~median) for signals with S/N $>$ 12.
Uncertainties in distances toward the Virgo cluster can affect the calculated
HI mass distribution.

\end{abstract}


\section {Introduction}\label{intro}

A census of HI in the local Universe 
will shed light on galaxy evolution and a census
of the gas content of galaxies at $z\sim 0$.  As such, a complete enumeration
tracing the HI at low redshift will yield a homogeneous dataset with 
quantitative measurements 
for kinematics, redshifts, and HI masses.  An HI catalog provided
by a wide-area survey serves as an outstanding complement to existing and future surveys
at other wavelengths, including the 
large area Two Micron All Sky Survey (2MASS; Skrutskie \etal 2006)
and the Sloan Digital Sky Survey (SDSS; York \etal 2000).  A high sensitivity HI survey 
sampling a large area of sky completes the picture by adding
information about the gas content of galaxies in the local Universe.

First generation HI surveys have been limited by sensitivity or areal coverage.
The HI Parkes All-Sky Survey (HIPASS; Barnes \etal 2001) has detected 5317 extragalactic sources
over 30,000 deg$^{2}$ of sky (Meyer \etal 2004; Wong \etal 2006) using a 
13 element receiver system.  The survey's 
successful catalogs provided a wide area of sky coverage.  The Arecibo Dual Beam Survey 
(ADBS; Rosenberg \& Schneider 2002)
surveyed a smaller area of sky, but utilized the advantage of the higher
sensitivity, smaller beam size, and better
positional accuracy of the Arecibo Radio Telescope.    The ADBS in particular showed the populations of objects
that could be detected with high sensitivity HI surveys that 
were not subjected to the biases of optical surveys.
A ``2nd generation'' extragalactic HI survey can combine high sensitivity with coverage of
a wide area and will yield a cosmologically fair sampling of HI 
sources in the volume of the local Universe.  
As astronomy pushes into a realm of large scale data mining, optical and IR surveys will be greatly 
complemented by the catalogs from HI surveys.

The Arecibo Legacy Fast ALFA (ALFALFA; Giovanelli \etal 2005a: Paper I) survey, currently underway,
utilizes the 7-element Arecibo L-band Feed Array (ALFA) receiver system to conduct a wide area
extragalactic HI investigation.  Using the high sensitivity of the 305 meter reflector, ALFALFA
will survey 7000 deg$^{2}$ of sky at high Galactic latitude up to $cz_{\odot}\sim$18000 \kms.
As a second generation HI survey, ALFALFA
improves upon previous surveys with larger bandwidth and finer spectral resolution,
as well as with higher sensitivity and angular resolution.  While the eight-fold
increase in sensitivity will assist in the detection of low HI mass galaxies, 
the wider bandwidth will allow ALFALFA to sample a fair volume of the Universe.
Higher angular resolution will largely avoid the source confusion that affected identification of
optical counterparts in surveys such as HIPASS.
Paper I describes the survey's science goals and objectives.  Based on simulations
and ALFALFA's initial catalogs (Giovanelli \etal 2005b, 2007; Saintonge \etal 2008),
more than 20,000 sources will be detected out to $z \sim 0.06$.

Once an HI mass is detectable at an astrophysically significant distance, maximizing
the survey area is more advantageous in increasing the detection rate
than increasing integration time.
The ``fast'' element of ALFALFA aims to survey a large amount of sky quickly, with
a 1-Hz sampling rate for scans, and 40 second integration time per pixel.
These observing parameters allow ALFALFA to detect
low HI mass galaxies down to M$_{HI} \sim 10^{5}$\msun~  in the Local Group and M$_{HI} \sim 2 \times 10^{6}$\msun~
at the distance of the Virgo Cluster.
As such, a crucial goal of using ALFALFA's 
low mass sample is a determination of the HI mass function (HIMF) both globally and in a variety of
environments.  Obtaining a statistically significant sample coupled with accurate
distances for nearby galaxies will yield an accurate 
and robust determination of the faint end slope of the 
HI mass function, one of the many science goals of ALFALFA and HI astrophysics.
Determining the statistics of these low-mass gas rich halos will also shed light on
the ``substructure'' problem (Somerville 2002), which predicts that we should observe more satellite objects
in the hierarchical structure formation paradigm.

ALFALFA is detecting a variety of objects including large disk galaxies and small low mass dwarfs
in the Local Universe.  Datasets provided by HI surveys like ALFALFA give systemic redshift, integrated
line fluxes from spectral profiles, and velocity width measurements for cataloged objects.
Approximately 500 galaxies will be resolved nearby.  In addition to single galaxies, ALFALFA also provides data
for extended tidal features and nearby high--velocity clouds, and gas-rich halos
which may exhibit little or no stellar component.

The survey's observing strategy has yielded an open shutter time of 97\%.  Final
data cubes are constructed after two-pass transit observations of targeted areas are completed.  ALFALFA's public website is
continually updated with schedules, survey information, and links to online catalogs.  The survey's
first catalogs are being published in periodic installments as target areas and data cubes are completed.
The first catalog (Giovanelli \etal~2007)
detailed the results of 730 detections from the Northern Virgo region, while the second (Saintonge \etal~2008)
yielded 488 detections in a direction opposite of the Virgo cluster.  The results presented 
in this paper extend the release of the ALFALFA survey with \ncat~ detections in the Virgo Cluster region
south of first catalog.

Virgo is the nearest rich-cluster of galaxies with over 1300 reported members (Binggeli, Sandage \& Tammann 1985; BST).
Because of its proximity, high density, and location in the sky visible
from Arecibo, Virgo is an important primary target area for ALFALFA.  
In addition, Virgo is often included in the coverage of both medium and 
large sky surveys (see ACS Virgo in C{\^o}t{\'e} \etal~2004; BST; HIPASS; SDSS), creating
an important synergy with one of ALFALFA's primary target areas.  Many late-type spiral galaxies
are found in the outlying periphery, which has not yet virialized.  As such, many of these gas rich galaxies
are still infalling into the cluster, in contrast to the Virgo core spirals, known to be highly 
deficient in HI (Davies \& Lewis 1973; Chamaraux \etal~1980; Giovanelli \& Haynes 1983; Solanes
\etal~2002).  The catalog presented in this paper consists of HI sources extracted in Right Ascension 
between $11^h36^m <$ R.A.(J2000) $< 13^h52^m$ and Declination between $+08^{\circ} <$ Dec.(J2000) $< +12^{\circ}$.
This area, covering 132 deg$^{2}$, includes the portion of Virgo between cluster 
members M87 and M49 and complements the earlier catalog by Giovanelli \etal (2007).

The paper organization is as follows.  Section 2 describes the observations and data reduction.  
Section 3 details the catalog contents and signal extraction reliability.
Section 4 describes the sample statistics of the dataset.
Section 5 deals with the positional accuracy of ALFALFA sources.
Section 6 summarizes the results.

\section{Observations and Data Reduction}\label{obs}

ALFALFA utilizes a fixed azimuth drift mode observation scheme with the 7-element
ALFA multi-beam receiver system.  Each dual polarization feed has a beam that
is $3.3$\arcmin~$\times$~ $3.8$\arcmin~in size.  Beam maps
of each feed can be found in Paper I.  The ALFA array is
rotated so that parallel feed tracks
of constant J2000 declination are spaced at 1.05\arcmin.  Scans are obtained with the telescope
parked along the meridian--small corrections in zenith angle are applied
between scans such that the beams are tracking along the constant epoch
J2000 Declination.  Fourteen simultaneous spectra are obtained at a sampling rate of 1 Hz 
in R.A. and
scans are composed of 600 one--second records each sweeping 10 minutes of Right Ascension.  The backend correlator setup uses
a 100 MHz bandwidth centered at 1385 MHz.  Raw scans have 4096 channels, giving
a spectral resolution of 24.4 kHz ($\delta v\sim 5.2$~\kms~ at $cz\sim$0).
Other survey details about the observing and
technical modes can be found in Paper I.

Raw data scans are processed offline for each observing session.  
Initial processing by team members includes flagging of radio interference,
bandpass and flux calibration, data quality assessment, and identifying 
strong continuum sources.  Observing runs
are scheduled to fill in target area tiles that are
composed of $2.4^{\circ} \times 2.4^{\circ}$ areas of sky. Once scheduled target area tiles are completed, the
level I data scans are combined into regularly gridded 3-D data cubes, each
2.4 $\times$ 2.4 degrees in spatial extent, and 4~$\times$~1024 channels in velocity space.
These data cubes contain spectral data header information, coordinates,
both polarizations, continuum data, map weightings, and scan makeup.
An automated matched filter algorithm is used for signal extraction (Saintonge 2007).
Candidate signals are later examined by eye, and integrated spectral profiles are 
created.  Flux measurements, velocities, and widths are cataloged for publication
and noted as possible followup candidates.
  The resulting data cubes are approximately 4~$\times$~380 MB in size.

\subsection{Continuum maps}

In addition to the spectral line data, continuum maps are also created
using data from the level I data scans, described in Giovanelli \etal (2007).
Channels are flagged through the bandpass calibration process and/or
the manual flagging process mentioned previously.
A background total power continuum value is also computed for all time series records for all
records and channels that have not been flagged, excluding point sources; the continuum
contribution from these point sources is also stored for the creation of the continuum maps.
Sources are detected with an automated
peak finding algorithm, which then searches a database of NVSS (Condon \etal~1998) sources
and looks for matches, comparing both fluxes and positions.  Final positions
are fit with a 2D Gaussian and stored along with peak flux measurements.
These measurements are used to correct for positions offsets as described in section 5,
and continuum source variability studies.

\subsection{Data Access}

As the ALFALFA collaboration includes many worldwide members,
rapid dissemination of the data
products is of the utmost importance to the survey's success.
The catalog results presented in this paper will be 
added to the existing ALFALFA archive at
{\it http://arecibo.tc.cornell.edu/hiarchive/alfalfa}/.
The site provides web services using protocols from the  U.S. National
Virtual Observatory\footnote{This research has made use of data obtained 
from or software provided by
the US National Virtual Observatory(NVO), which is sponsored by the National
Science Foundation.}.  The measurements
and spectra will join the fast growing archive of HI measurements
(Springob \etal~2005; Giovanelli \etal~2007; Saintonge \etal~2008).
An ongoing development effort focuses on the
long--term public delivery of the 3--D ALFALFA data set through web--based 
access tools. At this time, delivery of the 3-D data is made possible 
through the observing team itself, by direct contact to R.G. or M.P.H. 
A major challenge is data volume: after regridding the 3-D cubes covering the 34 individual 
``grids'' from which the current catalog each occupy 50 GB. 
Allowing access and manipulation of the gridded data publicly will require
the development of web services and server applications.

In this work, we present a catalog of HI sources extracted
from the ALFALFA grids covering a region stretching
from $+8^\circ$ to $12^\circ$ in Dec. and from 
$11^h36^m$ to $13^h52^m$ in R.A. For reference to our database,
denomination of the grids from which sources in this catalog were
extracted are 1140+09 to 1348+09 and 1140+11 to 1348+11, in
steps of 8$^m$ in R.A.
The solid angle subtended by this region is $\sim$132 \sqd, 
which is $\sim1.9$\% of the sky to be ultimately surveyed by
ALFALFA. The coverage of the region is complete by the target goals
of ALFALFA, i.e. the region has been sampled by two separate passes
with the ALFA array in drift mode. 

\section{Second Virgo Catalog}\label{sourcecat}

Raw data scans are processed to Level I (see Paper I for details) datasets upon completion
of the observing sessions.  The survey
employs a two--pass system to yield a well spatially
sampled dataset.  Once the survey has completed two passes
of a designated tile it is converted into regularly
gridded data cubes.  
Base-lining, signal extraction, and source identification are completed and
measurements are taken from integrated spectral profiles.  These profiles
originate from cutouts of the sub-grids, which are then used to create 
properly weighted source spectra.

Signal extraction is initiated in the Fourier domain with an automated matched
filter algorithm (Saintonge 2007).  Followup by eye confirms these sources and matches
them with any possible optical counterpart found in the Sloan Digital
Sky Survey (SDSS) or blue band imaging provided by the Digital Sky Survey.
In addition, measurements and sources can be compared to NED and 
listings of existing HI observations.  This facilitates
quick response for followup with selected sources.   Candidate detections are assigned
to the following classes:

1.  Signals that are extragalactic HI sources above a soft limit of S/N~$\gtrsim$~6.5

2.  Signals that are likely galactic or perigalactic in nature - high--velocity clouds.

3.  Signals of lower S/N ($4.5 \lesssim$~S/N~$\lesssim 6.5$) but have 
      corroborating evidence based on identification of an
      optical counterpart with similar redshift measurement at another wavelength.\\

Other low S/N candidates are placed in a separate listing for future followup.

Table 1 presents the results of the \ncat~ detections from this portion of the survey.
The column descriptions are as follows:

Col. 1: The catalog and source number

Col. 2: Arecibo General Catalog number.  This number corresponds to a private database
        entry maintained by R.G. and M.P.H.

Col. 3: Centroid position of the HI source (R.A. and Dec. J2000).  The listed position has been corrected
        for systemic pointing errors.  Positional accuracy will be discussed in section 5.

Col  4: J2000 position of the most likely optical counterpart of the HI detection.  The listed
        position has been examined by eye using SDSS or the Digital Sky Survey.
        The centroid accuracy is $\lesssim$2\arcsec.  Optical images are examined
        for counterparts based on spatial proximity, morphology, color, and redshift.
        If no optical counterpart can be clearly identified, no optical position is recorded in the catalog -- high--velocity
        cloud detections are included in this category.  Comments are provided 
        in column 14 if
        a preferable source is identified and other possible candidates are present
        in the field.

Col. 5: heliocentric velocity of the HI source in \kms, $cz_{\odot}$, measured 
	as the midpoint 
	between the channels at which the flux density drops to 50\% of 
	each of the two peaks (or of one, if only one is present) at each
	side of the spectral feature.  The error on $cz_\odot$
	to be adopted is half the error on the width, tabulated in Col. 7.

Col. 6: velocity width of the source line profile, $W50$, measured at the 50\%
	level of each of the two peaks, as described for Col. 5. This value 
	is corrected for instrumental broadening. No corrections due to
	turbulent motions, disk inclination or cosmological effects are
	applied.  The instrumental correction used is that described by Springob \etal (2005).
        Therefore, the expression for $W50$ is

        \be
	    W50=\sqrt{W50_{uncorr}^2-(2\Delta\nu)^2}
	\ee

	where $\Delta\nu$ is the channel separation in \kms.

Col. 7: estimated error on the velocity width, $\epsilon_w$, in \kms.
	This error is the sum in quadrature of two components: the first is a
	statistical error, principally dependent on the S/N
	ratio of the feature measured; the second is a systematic error
	associated with the subjective guess with which the observer estimates 
	the spectral boundaries of the feature: maximum and minimum guesses of 
	the spectral extent of the feature are flagged and the
	ratio of those values is used to estimate systematic errors on the
	width, the velocity and the flux integral. In the majority of cases,
	the systematic error is significantly smaller than the statistical
	error; thus the former is ignored.

Col. 8: integrated flux density of the source, $F_c$, in Jy \kms. This 
	is measured on the integrated spectrum, obtained by
	spatially integrating the source image over a solid angle of at
        least $7$\arcmin $\times 7$\arcmin ~and dividing by the sum of the survey beam
	values over the same set of image pixels (Shostak \& Allen 1980). 
	Estimates of integrated fluxes for very extended sources with
	significant angular asymmetries can be misestimated by our 
	algorithm, which is optimized for measuring sources comparable with
	or smaller than the survey beam. A special catalog with parameters
	of extended sources will be produced after completion of the survey.

Col. 9: estimated uncertainty of the integrated flux density, in Jy \kms .
	Uncertainties
	associated with the quality of the baseline fitting are not included;
	an analysis of that contribution to the error will be presented
	elsewhere for the full survey. See the description of Col. 7 for the
	contribution of a possible systematic measurement error.

Col. 10: signal--to--noise ratio S/N of the detection, estimated as 
        \be
        S/N=\Bigl({1000 F_c\over W50}\Bigr){w^{1/2}_{smo}\over \sigma_{rms}}
        \ee
        where $F_c$ is the integrated flux density in Jy \kms, as listed in  col. 8,
        the ratio $1000 F_c/W50$ is the mean flux across the feature in mJy,
        $w_{smo}$ is either $W50/(2\times 10)$ for $W50<400$ \kms or
        $400/(2\times 10)=20$ for $W50 \geq 400$ \kms [$w_{smo}$ is a
        smoothing width expressed as the number of spectral resolution
        bins of 10 \kms bridging half of the signal width], and $\sigma_{rms}$
        is the r.m.s noise figure across the spectrum measured in mJy at 10
	\kms resolution, as tabulated in Col. 11. In a similar analysis, in 
        Giovanelli \etal (~2005b; hereafter Paper II) we adopted a maximum smoothing 
	width $W50/20=10$. See Figure 4 ~and related text
	below for details. The value of the smoothing width could be
	interpreted as an indication of the degree to which spectral smoothing 
	aids in the visual detection of broad signals, against broad--band 
	spectral instabilities. The ALFALFA
	data quality appears to warrant a more optimistic adoption of
	the smoothing width than previously anticipated. 

Col. 11: noise figure of the spatially integrated spectral profile, $\sigma_{rms}$,
	in mJy. The noise figure as tabulated is the r.m.s. as measured over the signal-- and
	RFI-free portions of the spectrum, after Hanning smoothing to a spectral
	resolution of 10 \kms.

Col. 12: adopted distance in Mpc, $D_{Mpc}$. For objects with $cz_{cmb}> 3000$ \kms, 
	the distance is simply  $cz_{cmb}/H_\circ$; $cz_{cmb}$ is the recessional velocity
	measured in the Cosmic Microwave Background reference frame and $H_\circ$ is
	the Hubble constant,  for which we use a value of 72 \kms Mpc$^{-1}$.
	For objects of lower $cz_{cmb}$, we use a peculiar velocity model for the
	local Universe, as described in Paper II. Objects which are thought to be parts
	of clusters or groups are assigned the $cz_{cmb}$ of the cluster or group.
	Cluster and group membership are assigned following the method described
        in Springob \etal (2007). A detailed
	analysis of group and cluster membership of ALFALFA objects will be presented
	elsewhere. Note that the Virgo cluster extends over much of the solid angle 
	surveyed. This introduces unavoidable ambiguities in the distance assignment,
	as the peculiar flow model only corrects for large--scale perturbations in the
	velocity field and is unable to deal effectively with regions in the
	immediate vicinity of the cluster and along a section of a cone which contains
	the cluster, up to $cz\sim 2500$ \kms.

Col. 13: logarithm in base 10 of the HI mass, in solar units. This parameter is 
	obtained by using the expression $M_{HI}=2.356\times 10^5 D_{Mpc}^2 F_c$. 

Col. 14: object code, defined as follows: 
	
	Code 1 refers to sources 
	of S/N and general qualities that make it a reliable detection.
	By ``general qualities'' we mean that, in addition to an approximate
	S/N threshold of 6.5, the signal should  exhibit a good match between
	the two independent polarizations and a spatial extent consistent
	with expectations given the telescope beam characteristics. Thus, some
	candidate detections with $S/N>6.5$ have been excluded on grounds
	of polarization mismatch, spectral vicinity to RFI features or peculiar
	spatial properties. Likewise, some features of $S/N<6.5$ are included
	as reliable detections, due to optimal overall characteristics of
	the feature. The S/N threshold for acceptance of a reliable detection
	candidate is thus soft. In a preliminary fashion, we estimate that
	detection candidates with $S/N<6.5$ in Table 1 are reliable, i.e. they
	will be confirmed in follow--up observations in better than 95\% of
	cases (Saintonge 2007). 

	Code 2 refers to sources of low S/N ($<$ 6.5), which would  
        ordinarily not be considered
	reliable detections by the criteria set for code 1. However, those
	HI candidate sources are matched with optical counterparts with known
	optical redshifts which match those measured in the HI line. These
	candidate sources, albeit ``detected'' by our signal finding algorithm,
	are accepted as likely counterparts only because of the existence of
	previously available, corroborating optical spectroscopy. We refer to
	these sources as ``priors''. We include them in our catalog because
	they are very likely to be real.

	Code 9 refers to objects assumed to be HVCs; no
	estimate of their distances is made.

 	Notes flag. An asterisk in this column indicates a comment is included
	for this source in the text below.

Only the first few entries of Table 1 are listed in the printed version of this
paper. The full content of Table 1 is accessible through the electronic version
of the paper and will be made available also through our public digital 
archive site.

\begin{deluxetable}{cccccccccccc}
\rotate
\tablewidth{0pt}
\tabletypesize{\scriptsize}
\tablecaption{HI Candidate Detections\label{alfadet}}
\tablehead{
\colhead{Cat\#-ID} &\colhead{AGC}   & \colhead{HI Coords (2000)} & \colhead{Opt. Coords.} &
\colhead{cz$_\odot$}  & \colhead{$w50 ~(\epsilon_w$)} &
\colhead{$F_{c} ~(\epsilon_{f})$} & \colhead{S/N} & \colhead{rms} &  
\colhead{Dist}    & \colhead{$\log M_{HI}$} & \colhead{Code} 
    \\
 & & & & {\kms} & {\kms} & {Jy \kms} & & {mJy} & Mpc & {$M_\odot$} &  
}
\startdata
3-  1  & 210497 &  113610.7+100304 &  113609.9+100319 &  6210 &  261 (23) & 2.58 (0.10) &  17.1 &  2.09 &  88.2 &  9.67 & 1     \\
3-  2  & 213318 &  113610.9+114858 &  113612.2+114857 &  9518 &  229 (13) & 0.68 (0.06) &   6.2 &  1.61 & 133.7 &  9.46 & 2     \\
3-  3  & 210517 &  113653.6+115040 &  113655.3+115053 & 10347 &  297 ( 5) & 1.16 (0.07) &   9.3 &  1.62 & 145.2 &  9.76 & 1     \\
3-  4  & 213320 &  113713.2+114840 &  113714.3+114806 & 10714 &  122 ( 8) & 1.30 (0.05) &  17.6 &  1.49 & 150.3 &  9.84 & 1     \\
3-  5  & 213100 &  113739.2+085134 &  113739.2+085151 &  3815 &  118 (36) & 0.50 (0.06) &   5.4 &  1.91 &  55.4 &  8.56 & 2     \\
3-  6  & 210540 &  113805.9+111243 &  113808.1+111149 & 10660 &  316 (15) & 1.76 (0.10) &  10.5 &  2.11 & 149.5 &  9.96 & 1     \\
3-  7  &   6605 &  113813.3+120646 &  113813.0+120643 &   982 &   89 ( 6) & 3.20 (0.07) &  35.7 &  2.11 &  11.6 &  8.00 & 1     \\
3-  8  & 215419 &  113830.7+111357 &  113832.2+111316 & 13062 &   53 (14) & 0.63 (0.06) &   8.1 &  2.37 & 182.8 &  9.70 & 1     \\
3-  9  & 210550 &  113844.5+105327 &  113846.1+105258 & 12788 &  301 (36) & 1.21 (0.12) &   5.7 &  2.73 & 179.0 &  9.96 & 2     \\
3- 10  & 213025 &  113905.0+100815 &  113905.4+100807 & 12721 &  118 ( 5) & 1.14 (0.07) &  11.0 &  2.11 & 178.1 &  9.93 & 1     \\
3- 11  & 213102 &  113911.8+093815 &  113911.6+093808 &  5963 &  137 (15) & 0.93 (0.07) &   9.0 &  1.96 &  84.8 &  9.20 & 1     \\
3- 12  & 213103 &  113924.3+084254 &  113926.3+084222 &  5460 &  180 ( 6) & 0.76 (0.07) &   6.1 &  2.10 &  77.9 &  9.04 & 2     \\
3- 13  &   6626 &  113952.1+085214 &  113952.8+085229 &  1984 &  156 (10) & 4.87 (0.07) &  43.1 &  2.02 &  29.6 &  9.00 & 1     \\
3- 14  & 210582 &  114000.7+112745 &  114001.5+112746 & 10267 &  360 ( 4) & 2.86 (0.10) &  15.6 &  2.16 & 144.1 & 10.15 & 1     \\
3- 15  &   6633 &  114018.8+090033 &  114018.5+090035 &  1810 &  294 (24) & 4.12 (0.10) &  23.4 &  2.29 &  29.6 &  8.93 & 1     \\
\hline
\enddata
\end{deluxetable}

\subsection{Notes and Unique Detections}

In addition to the detections reported by Kent \etal (2007) and Haynes \etal (2007) that exhibit 
no apparent optical counterparts, a number of other detections exhibit unique properties:  
ambiguous optical identification in need of optical followup, disturbed HI morphologies, 
and HI connecting members of a galaxy pair or group. Some have presented
challenges in the data reduction process, and special care has been 
taken to accurately extract the spectra.  Other clear detections may or may not
be associated with previously uncataloged low surface brightness detections
in the vicinity of a given HI centroid.
Many of these detections will be described in future publications
with the added benefit of followup observations and detailed analysis
both at radio and other wavelength regimes (Kent \etal in preparation).\\
	
\noindent Notes associated with the objects listed in Table 1 follow:

{\footnotesize

\noindent 3-  6      In group with AGC 213325 and AGC 213324

\noindent 3-  8      Crowded opt field, ambiguous opt id

\noindent 3- 26      Blend with UGC 6668/NGC 3825; part of WBL 350-005;  10 members in grp    
 
\noindent 3- 27      Blend with UGC 6661/NGC 3822   

\noindent 3- 29      Poor data quality: params uncertain

\noindent 3- 43      Blend with UGC 6692/NGC 3833

\noindent 3- 48      Poor data quality; params uncertain

\noindent 3- 50      Edge of single pass; params uncertain

\noindent 3- 52      Possible blend with UGC 6715

\noindent 3- 53      HI merges in spectral region affected by rfi; params very uncertain 
   
\noindent 3- 56      Very lsb optical counterpart

\noindent 3- 60      Single pass coverage; params very uncertain 
 
\noindent 3- 61      Extension of HI 2\arcmin~ to the north.

\noindent 3- 70      Ambiguous optical id; could be object to  at 114603.7+105127
 
\noindent 3- 79      Poor data quality: params uncertain

\noindent 3- 84      Possible HI envelope extended to the S.

\noindent 3- 93      Only one drift contributing.

\noindent 3- 95      Blend with HI tail leading N UGC 6871 

\noindent 3-115      Ambiguous optical id; also possible AGC 211853 at 115950.9+084735

\noindent 3-116      Poor data quality: params uncertain  

\noindent 3-117      Poor baseline: params uncertain

\noindent 3-125      Emission merges in region affected by rfi: params uncertain

\noindent 3-130      Single pass only; params uncertain

\noindent 3-136      Extended HI tail leading 4\arcmin south

\noindent 3-145      In group with UGC 7066
    
\noindent 3-147      In group with UGC 7066 and AGC 224237 at 120447.1+103735

\noindent 3-149      Interacting system with AGC 220075 at 120536.4+085917

\noindent 3-152      Emission merges in region affected by rfi:  params very uncertain.
  
\noindent 3-153      Params uncertain

\noindent 3-155      No discernible optical counterpart (see Kent \etal 2007)

\noindent 3-157      Ambiguous optical id: AGC 224552 also possible to N at 120916.80+103500

\noindent 3-161      Galaxy emission merges with region affected by rfi: params uncertain; opt id ambiguous: other possible counterpart AGC 224603 at 121018.81+112105

\noindent 3-165      Poor baseline: params uncertain

\noindent 3-182      Large z mismatch between opt and HI; no other clear opt counterpart for HI

\noindent 3-187      Very LSB counterpart.

\noindent 3-188      HI tail extending 5\arcmin NE.

\noindent 3-189      Very LSB counterpart.

\noindent 3-191      Triple peaked spectrum

\noindent 3-194      Very LSB counterpart.

\noindent 3-201      Ragged data; params uncertain

\noindent 3-208      Possible optical id also at 121843.14+114334  

\noindent 3-216      Pair with UGC 7383 
  
\noindent 3-218      Ambiguous optical id: also possible at 122026.5+085023


\noindent 3-221      Ambiguous optical id: also possible at AGC 220015 at 122040.8+083538

\noindent 3-222      Ambiguous optical id: also possible at AGC 224621 at 122045.9+115340

\noindent 3-227      Blend with UGC 7414
   
\noindent 3-228      Blend with AGC 7407
 
\noindent 3-229      near Galactic MW HI

\noindent 3-236      Ambiguous optical id: also possible at 122212.4+115515

\noindent 3-241      Emission merges in region affected by rfi:  Params very uncertain.

\noindent 3-255      Positional offset significant; optical id uncertain: also possible at 122609.9+080949

\noindent 3-258      Blend with AGC 224604 at 122629.6+090109
 
\noindent 3-259      Pair with UGC 7546

\noindent 3-260      Pair with UGC 7537

\noindent 3-266      Extended HI tail

\noindent 3-278      Very LSB counterpart

\noindent 3-282      Poor baseline; params uncertain

\noindent 3-285      No discernible opt counterpart.  Discussed in Kent \etal (2007).

\noindent 3-287      Detected by Oosterloo \& Van Gorkom and also McNamara \etal  Discussed in Kent \etal (2007).


\noindent 3-289      Poor baseline; params very uncertain

\noindent 3-290      No discernible optical counterpart.  Discussed in Kent \etal (2007).

\noindent 3-291      No discernible optical counterpart.  Discussed in Kent \etal (2007).  
  
\noindent 3-292      VCC1295; LSB counterpart.

\noindent 3-294      Possible opt id with v LSB object 1.5' to NE; positional offset signficant 

\noindent 3-295      Poor baseline - vicinity of M87.     

\noindent 3-296      No discernible optical counterpart.  Discussed in Kent \etal (2007).

\noindent 3-299      Very LSB optical counterpart

\noindent 3-300      Poor baseline: Close proximity to continuum source; params uncertain.

\noindent 3-303      Affected by proximity to MW emission; params. mildly uncertain.

\noindent 3-307      No optical counterpart; extended HI: HVC

\noindent 3-311      No optical counterpart; extended HI: HVC

\noindent 3-315      No optical counterpart; extended HI: HVC

\noindent 3-323      Blend of UGC 7777/7776, separation of pair is 2\arcmin.

\noindent 3-324      Emission affected by rfi:  Params mildly uncertain.

\noindent 3-341      Confused blend with galaxies AGC 226457 at 124020.7+081221 and AGC 224170 at 124023.0+081034

\noindent 3-342      Emission merges into region affected by rfi: params uncertain

\noindent 3-347      Affected by rfi; params mildly uncertain
   
\noindent 3-349      Ambiguous optical id; also possible counterparts at 124133.0+082110 and 124123.9+082158
         
\noindent 3-352      No optical counterpart; extended HI: HVC

\noindent 3-354      Affected by rfi; params mildly uncertain

\noindent 3-357      Part of a group;  possible extended envelope.

\noindent 3-359      Ambiguous opt id; also possible counterpart at 124351.3+110254

\noindent 3-361      Affected by rfi; params uncertain

\noindent 3-362      Affected by rfi; params uncertain

\noindent 3-364      Edge of map: params uncertain; optical id ambiguous: also possible counterpart AGC 220986 at 124445.6+094526

\noindent 3-365      Pair of galaxies; optical id ambiguous; also possible counterpart at 124456.7+101627
  
\noindent 3-366      Ambiguous opt id; also possible counterpart at 124507.8+094640
  
\noindent 3-371      No optical counterpart; extended HI: HVC 

\noindent 3-377      No optical counterpart; extended HI: HVC

\noindent 3-378      No optical counterpart; faint LSB at 124641.7+102309, 2.3\arcmin N 

\noindent 3-380      No optical counterpart; extended HI: HVC

 \noindent 3-386     Extended HI emission

\noindent 3-388      Optical id ambiguous: also possible AGC 221031 at 124835.5+090733

\noindent 3-392      Very LSB counterpart

\noindent 3-395      No optical counterpart; extended HI: HVC

\noindent 3-400      No optical counterpart; extended HI: HVC

\noindent 3-403      No optical counterpart; extended HI: HVC

\noindent 3-404      Emission from apparently blank opt field, in vicinity of UGC8037. See Kent \etal 2007

\noindent 3-405      Possible blend with UGC8045 at 3.2\arcmin~ E

\noindent 3-406      Emission from apparently blank opt field, in vicinity of UGC 8037.  See Kent \etal 2007

\noindent 3-407      Emission from apparently blank opt field, in vicinity of UGC 8037 and AGC 221110. See Kent \etal 2007

\noindent 3-408      Blend with emission of UGC 8042 at 1.5\arcmin~ SE

\noindent 3-410      No optical counterpart; extended HI: HVC

\noindent 3-411      Very LSB.

\noindent 3-413      No optical counterpart; extended HI: HVC

\noindent 3-420      Blend with emission of UGC 8089 at 1\arcmin~ NW

\noindent 3-455	 Emission merges into area affected by rfi

\noindent 3-458      Possible blend with emission from galaxy at 131021.6+084510, 2\arcmin~E
    
\noindent 3-461      Affected by rfi, params uncertain; ambiguous opt id: also possible id with AGC230140 at 131055.7+115230 or AGC230138 at 131050.8+115216

\noindent 3-469      Possible blend with emission from AGC 231046 at 131313.8+094052
     
\noindent 3-474      Possible blend with AGC 230192 at 131500.11+100142.6; $cz$ differs by 300~\kms.

\noindent 3-478      No optical counterpart; extended HI: HVC
     
\noindent 3-486      No optical counterpart; extended HI: HVC

\noindent 3-490      Blend with UGC 8421

\noindent 3-497      Very near the spectral band edge; params and detection uncertain

\noindent 3-498      Ragged data; params uncertain
   
\noindent 3-503      Possible blend with galaxy at 132850.5+114508

\noindent 3-522      Possible blend with AGC 230401 at 133218.8+102648

\noindent 3-543      Affected by rfi; params uncertain

\noindent 3-546      Affected by rfi; params very uncertain
   
\noindent 3-558      Affected by rfi; params very uncertain

\noindent 3-563      Extended HI emission in N-S direction
   
\noindent 3-568      Blended with AGC 230646 at 134653.8+113703
            
\noindent 3-570      Possible blend with AGC 233837 at 134728.5+095134

}

\section{Statistics of the Catalog}\label{stats}

\ncat~detections are presented for this region: \ndet ~($\sim$76\%) 
are code ``1'' detections that rate fair to excellent in quality;
\nprior ~(23\%) are code ``2'' detections that have known priors
associated with them -- optical redshifts corroborate these lower
S/N detections; the remaining \nhvc ~detections ~(2\%) are
high--velocity clouds.  The completeness and reliability
of the ALFALFA detection process are discussed by Saintonge (2007).
Follow-up observations on lower signal-to-noise candidates from the entire
survey are currently underway and will be discussed in future publications.

26\% of the detections presented in this catalog have $cz_{\odot} < 3000$ \kms, 
a large fraction explained by the fact that this portion of the survey crosses
the Virgo region and supergalactic plane where the density of galaxies is higher.
The detection rate is 4.3 objects per square degree.
Excluding the HVC detections, $\sim$2\% do not have
apparent optical counterparts.  These ``optically unseen'' HI clouds
were reported by Kent \etal (2007).  Of particular interest is
a cloud complex of five detections located halfway between M87
and M49.  If not gravitationally bound, the system will disperse
in less than a cluster crossing time.  The detections
and aperture synthesis followup are described by Kent (2007b).

The described catalog, when combined with that of strips
to the North presented by Giovanelli \etal (2007) presents a wealth
of information to complement existing and future Virgo Cluster
surveys across the spectrum.  The survey's
power in redshift detection in the local Universe can 
be compared with previous HI surveys.  ALFALFA improves upon previous
HI surveys in numbers
and positional accuracy of the first generation
HI survey HIPASS.  
Wong \etal (2006)
reported on 40 detections 
from the HIPASS survey in the same ALFALFA region for this paper's catalog.
ALFALFA also complements the {\it GOLDMINE} database (Gavazzi \etal~2003),
an optically selected compilation of Virgo galaxies.  Long integrations
were used on some of the {\it GOLDMINE} detections, as a result
some are fainter than the ALFALFA detection limit.  
16\% of the detections presented in this catalog give
new redshifts, and 60\% are detected in HI for the first time.  

The region of sky presented with this catalog lies south of M87 and
 is shown in Figure 1.  The upper panel displays
all detections, while the lower panel shows detections found at
$cz_{\odot} < 3000$~\kms.  A R.A. versus $cz_{\odot}$ cone diagram is shown in Figure 2.  
Various distributions of the catalog parameters are shown in Figure 3:
(a) shows large scale structures of Virgo and the A1367-Coma supercluster regime in
a velocity distribution, as well as the high $cz$ gap near 15,000 \kms resulting from RFI; panels
(b), (c), and (d) show the distributions
for the velocity width W50, the integrated flux and S/N ratio respectively.
The final panel (e) shows the HI mass distribution, with distances computed
from velocity flow models (Tonry \etal 2000; Masters \etal 2004). The 
uncertainty in the distance of galaxies projected on a narrow cone centered on the Virgo Cluster causes detections to 
have uncertain calculated HI masses.  Other distance indicator techniques
will need to be employed in order to properly determine an accurate HI mass function.
This topic will be discussed in future publications.

The top panel of Figure 4 shows the Sp\"{a}nhauer diagram, detailing the HI mass vs. distance for 
the objects in the catalog.  The placement of nearby objects
at the Virgo distance by the velocity flow model is clearly seen as the vertical
gathering of objects at 16.7 Mpc.  Regions affected by RFI are also indicated with
vertical dashed lines.  The middle panel of Figure 4 shows the S/N ratio versus
measured velocity width of the catalog entries, demonstrating the independence of
signal extraction on velocity widths.  The bottom panel
of Figure 4 shows the integrated flux versus measured velocity width.  The dashed
line indicates the a S/N limit of 6.5, the rough limit of code 1 objects listed
in Table 1. 

\section {Positional Accuracy of Cataloged Sources \label{pos}}

Continuum sources are extracted and their positions measured in a ``telescope''
coordinate reference frame.  This reference frame is determined at the Observatory
separately for each receiver system by optimizing a model fit to a set of
pointing calibrators.  Repeatable residuals from this model translate into small
scale pointing errors, which typically have amplitudes on order of 15\arcsec~ for
the ALFA array receivers.  These errors vary with both azimuth and zenith angle.  
Because ALFALFA observations are made with ALFALFA at the meridian (constant azimuth), 
and consist of long runs at constant zenith angle, the ALFALFA sources' positions in 
the telescope reference frame easily reveal the systematic errors represented by those 
residuals, when they are compared with the positions of a 
higher resolution -- and thus higher accuracy -- coordinate reference system.  
In this section we compare HI source positions obtained in the Arecibo telescope 
reference system with two such, more accurate position sets:  those of optical 
galaxies and those of VLA radio continuum sources (NVSS; Condon \etal 1998).

\subsection{Positional Differences between HI and Optical Sources}

The accuracy of positional HI centroids is important in establishing correct optical
identifications.  The ALFALFA
catalogs are presented to the astronomical community with
positions as accurate as possible.  As this catalog lies just south of the Paper III catalog,
the telescope configuration used in obtaining the data is similar.  The impact
of the telescope pointing is similarly recovered as we describe below.

The 3.3\arcmin$\times$3.8\arcmin~ FWHM elliptical beam of the Arecibo telescope 
affords the ALFALFA
survey greater positional accuracy than previous large scale blind HI surveys.
This factor in the survey characteristics becomes important
when matching HI detections with their optical counterparts,
if they exist. Positional accuracy is limited by the beam 
size, signal--to--noise, and errors in the telescope's pointing.  For
the region of sky surveyed in this paper, the ALFA receiver system
is oriented such that the beam's major axis
is in the Declination direction and the telescope's
azimuth arm is aligned in the North--South direction.

Processed ALFALFA drift scans are combined and re-gridded
into a data cube that is spatially sampled at 1\arcmin$\times$1\arcmin~.
Gaussian weighting is also applied and the final resolution
is 3.8\arcmin$\times$4.3\arcmin.
Sources and their associated measured parameters are
extracted from these derived data cubes.  Source extraction
is carried out via an automated matched filter algorithm
with visual followup confirmation by a project team member.
Once a source is identified, a 2-D map is integrated over the spectral
extent of the object, with ellipses fit to the quarter and 
half--power levels of the measured peak, as well as other preset
isophotal levels.  As most sources are unresolved by the ALFA beam,
the HI source position is taken to be the center of the half--power
ellipse.  Extended objects that may 
exhibit warped or asymmetric disks are treated with extra care
as the ellipse may not provide the best estimate of the 
position center.

The systematic pointing errors of the telescope can be recovered
due to the fixed azimuth telescope observing mode employed by the survey.
These errors add to $\sim$15\arcsec~ or more at L--band frequencies with
the maximum pointing errors occurring when the telescope is at high (or very low)
 zenith angles.

Positions of optical counterparts of HI sources are obtained by team members through
the Digital Sky Survey, from the SDSS database, or through the local
Arecibo General Catalog maintained at Cornell.
Figure 5 shows the 
positional differences (HI minus optical) for four different
Declination ranges, each binned by 1$^{\circ}$.
The points correspond to sources that have optical counterparts --
optically inert HI clouds, HVC, and tidal tails are not included. 
The position differences can be attributed to centroiding statistical
errors, optical misidentification, and the previously mentioned
telescope pointing offsets.  The optical centroid errors are negligible when
compared to the other source errors.  The telescope pointing errors
contribute to the majority of the systematic error budget, as illustrated by
offsets in the data distribution from the center of the plots.

Signal--to--noise affects the accuracy of
the HI position centroid.  Figure 6 shows the positional
accuracy of the HI positions, after correction for telescope pointing errors, for various S/N regimes.
For the higher S/N detections the median centroid error is approximately 15\arcsec.

\subsection {Positional Errors as Derived from Radio Continuum Sources}

In addition to spectral data cubes, ALFALFA data
are used to generate regularly gridded continuum maps.
Continuum sources are extracted and their positions determined.
These maps are used to determine pointing corrections for the ALFALFA
catalog objects by comparing
centroid positions with those published by
the NRAO VLA Sky Survey (NVSS; Condon \etal 1998).  
The positions presented in this catalog
and in catalogs published on the online ALFALFA data archive are 
corrected via the method presented here.  The pointing
corrections utilize data from sixty-four data cubes from the Paper III catalog
and the catalog presented in the paper.

Continuum sources for position comparison are selected with a priori 
information from the NVSS.
For identification accuracy, 
the positions of high signal--to--noise objects
are used.
Therefore, for each region of sky defined by an ALFALFA data cube, an NVSS source
subset
is identified with a peak intensity $I_{NVSS} > 20$~mJy beam$^{-1}$.
The average r.m.s of the NVSS source catalog is 0.48 mJy beam$^{-1}$.
For each of the sources in the subset, a 9\arcmin~ box
is cutout for the ALFALFA data centered on the NVSS centroid position.
A 2-D Gaussian fit is attempted in this window.  If the fit converges,
then the position from peak of the fit is compared with any NVSS sources
in the box.  This process consists of comparing the ALFALFA continuum
centroid with any possible NVSS positions and flux totals in the 9\arcmin~ box.
If the sum of the NVSS fluxes differs from the measured ALFALFA flux
by more than 50\%, then the source is not used in the pointing correction.
Because of the differences in beam sizes,
several NVSS sources may be confused and seen
as one source in the ALFALFA beam.  Therefore, if 
flux agreement is achieved between the ALFALFA and NVSS entries, 
acceptable sources are saved and position differences are computed in Right
Ascension and Declination.

Using this automated method for cross-identification of NVSS sources yields
1583 objects with centroid positions in the ALFALFA data cubes covering 
$\sim$240 deg$^{2}$ of sky.  The differences between the ALFALFA
and NVSS centroids are plotted in Figure 7.  These position differences
are binned in quarter degree Declination increments.  
The robust mean position for each Declination bin is plotted in Figure 8,
with vertical bars indicating the error on the mean and the horizontal
bars indicating the bin width.  A third order polynomial is fit through
these averaged points, and the resulting coefficients are used to 
correct the pointing in the catalog.  The positions 
for the HI detections published in this
paper reflect these pointing corrections, which remove telescope pointing
errors from the HI centroids in our catalogs.

\section {Summary}\label{summary}

This paper completes another important part of the ALFALFA survey toward the
Virgo Cluster region.  The catalog presented here completes the Virgo ``core''
area centered around M87 and its periphery.  The sample of galaxies detected
yield a redshift distribution exhibiting large scale structure from Virgo
and the Coma Supercluster regime, shown in Figures 2, 3 and 4.
The catalog covers 132 \sqd~of sky south of M87, completing another 2\% of the survey.
The complementary conclusions to the Paper III catalog are as follows:

\begin{itemize}
\item
The third ALFALFA catalog installment presented here completes the core
of the Virgo cluster region initiated with Paper I.  Together, Paper I
and Paper VI complement the initial anti-Virgo catalog (Saintonge \etal 2008)
for future analysis.

\item
Objects of interest include a cloud complex lying midway between
M87 and M49 that do not correlate with any apparent optical counterpart.
The complex exhibits a velocity dispersion of $\sim$250~\kms~ between
400 and 640~\kms.  The spatial extent of the complex is over 200 kpc
if it is at the Virgo distance.   An 
analysis of this system, including aperture synthesis followup,
will be detailed in Kent \etal (in preparation) .

\item
The median $cz$ for the catalog presented here is 6500~\kms; the redshift distribution 
is analogous to that of the first Virgo catalog (Giovanelli \etal 2007).

\item
A number of high velocity clouds with velocities between -150 and +200 \kms~are
detected.  Given the proximity of the region surveyed to the North
Galactic Pole, their heliocentric velocities are minimally affected by 
galactic rotation and represent true infall or outflow motions.

\end{itemize}

In conclusion, we present the third catalog release of the ALFALFA survey,
bringing the total projected sky coverage to $\sim$5\%. 
This installment of the survey and its associated data products will be incorporated
into the more extensive digital HI dataset at
{\it http://arecibo.tc.cornell.edu/hiarchive/alfalfa/}.

RG and MPH acknowledge the partial support of NAIC as Visiting 
Scientists during the period of this work. RAK and TB thank NAIC for partial support. RAK, TB, and NB
are grateful to NAIC and the Cornell University Astronomy Department
for its hospitality to them as sabbatic visitors.  This work has been supported by NSF 
grants AST--0307661, AST--0435697, AST--0347929, AST--0407011, AST--0302049;
and by a Brinson Foundation grant. We thank the Director of
NAIC, Robert Brown, for stimulating the development of major ALFA surveys, 
H\' ector Hern\' andez for his attention to the telescope scheduling and the Director, 
telescope operators and support staff of the Arecibo Observatory for their proactive 
approach. We thank Tom Shannon for his advice and assistance with hardware, 
system and network issues at Cornell.  We also thank the referee for
their useful comments on this work.

\vskip 0.3in
 
\newpage

\begin{figure}[h]
\plotone{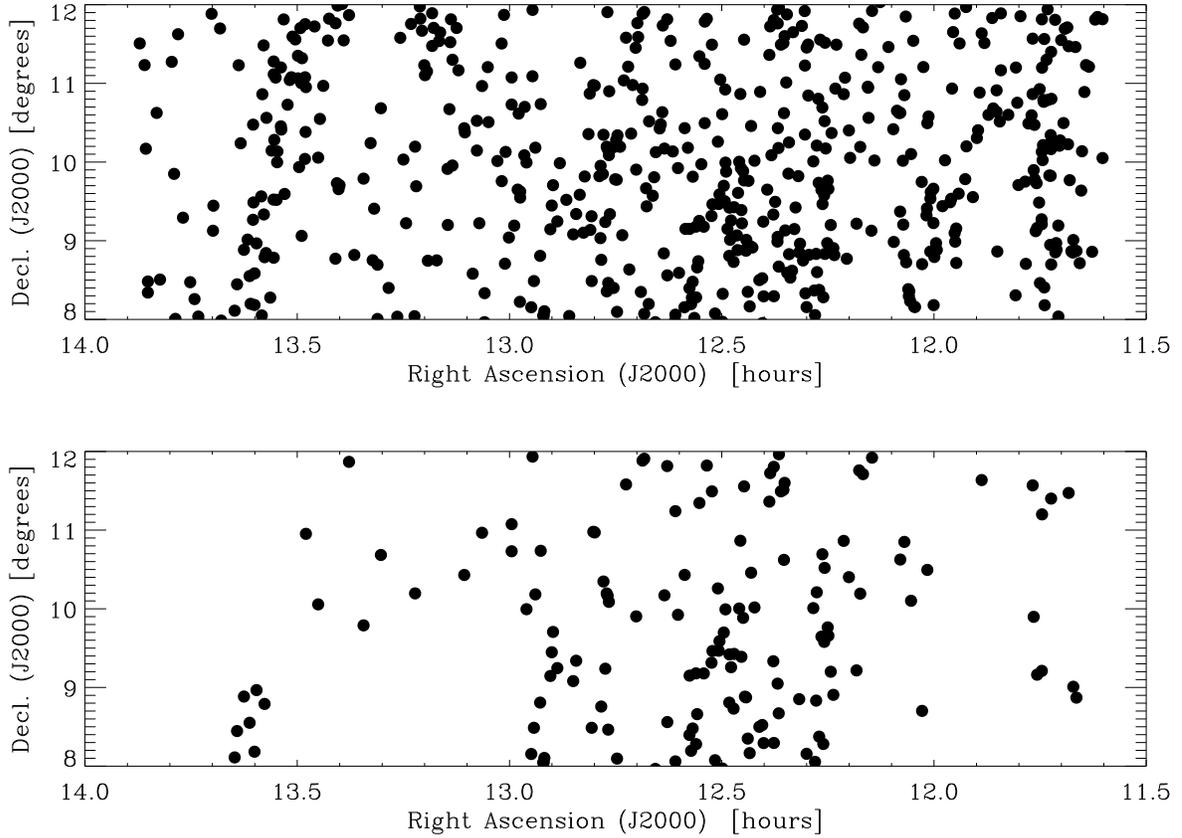}
\caption{Spatial distribution of HI candidate detections
listed in Table 1.  The upper panel shows all detections
presented in this catalog.  The lower panel shows
all objects with  $cz_{\odot}<3000$~\kms~.}
\label{skyplot}
\end{figure}

\begin{figure}[h]
\epsscale{0.7}
\plotone{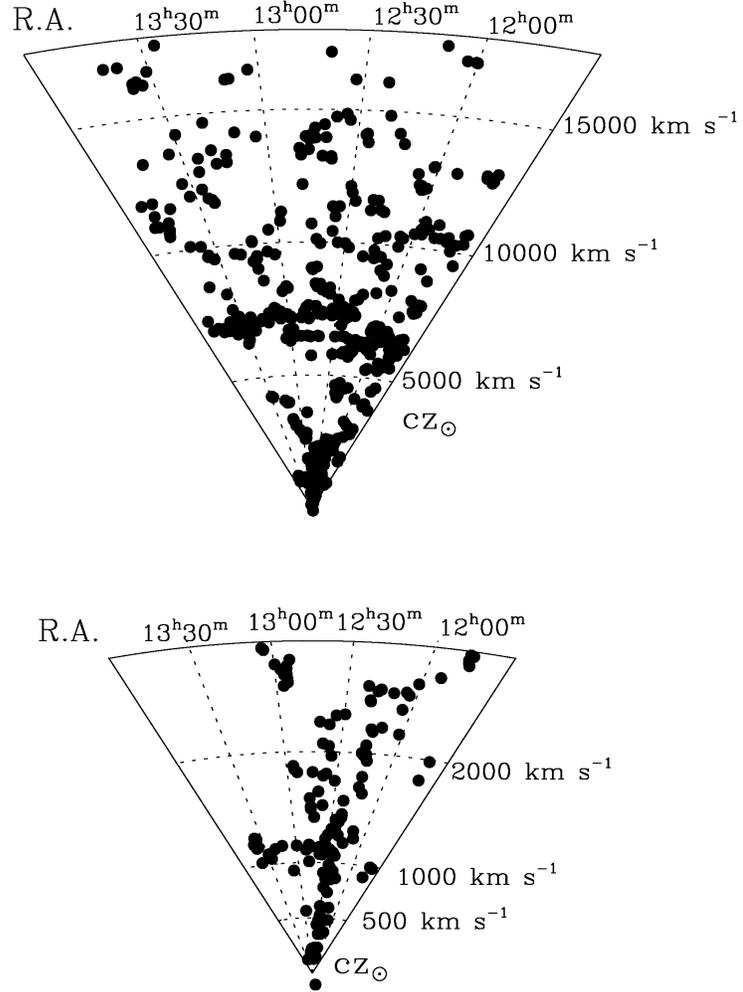}
\caption{Cone diagrams showing right ascension
vs. $cz_{\odot}$ for sources listed
in Table 1.  The top diagram shows all data for this paper's catalog.
The bottom diagram shows a close-up view for data at $cz_{\odot}~\leq~$3000~\kms.}
\label{coneplot}
\end{figure}

\begin{figure}[h]
\epsscale{0.5}
\plotone{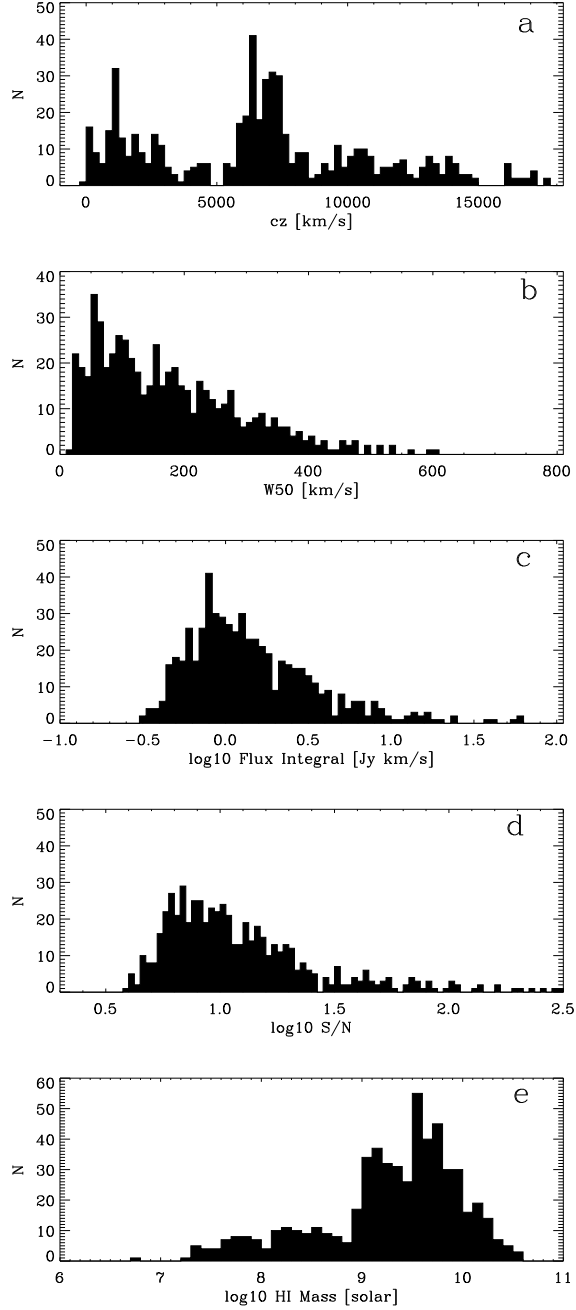}
\caption{Various histograms of the HI detections
listed in Table 1.  (a) shows the $cz$ distribution in \kms;
(b) shows the HI line width W$_{50}$ in \kms;
(c) shows the base 10 logarithm of the integrated flux in Jy \kms;
(d) shows the base 10 logarithm of the peak S/N ratio;
(e) shows the derived HI mass, in base 10 logarithmic units of \msun.  The
HI masses of some detections may be skewed to higher estimates
due to distance uncertainties toward the Virgo cluster region.}
\label{histograms}
\end{figure}

\begin{figure}[h]
\epsscale{0.8}
\plotone{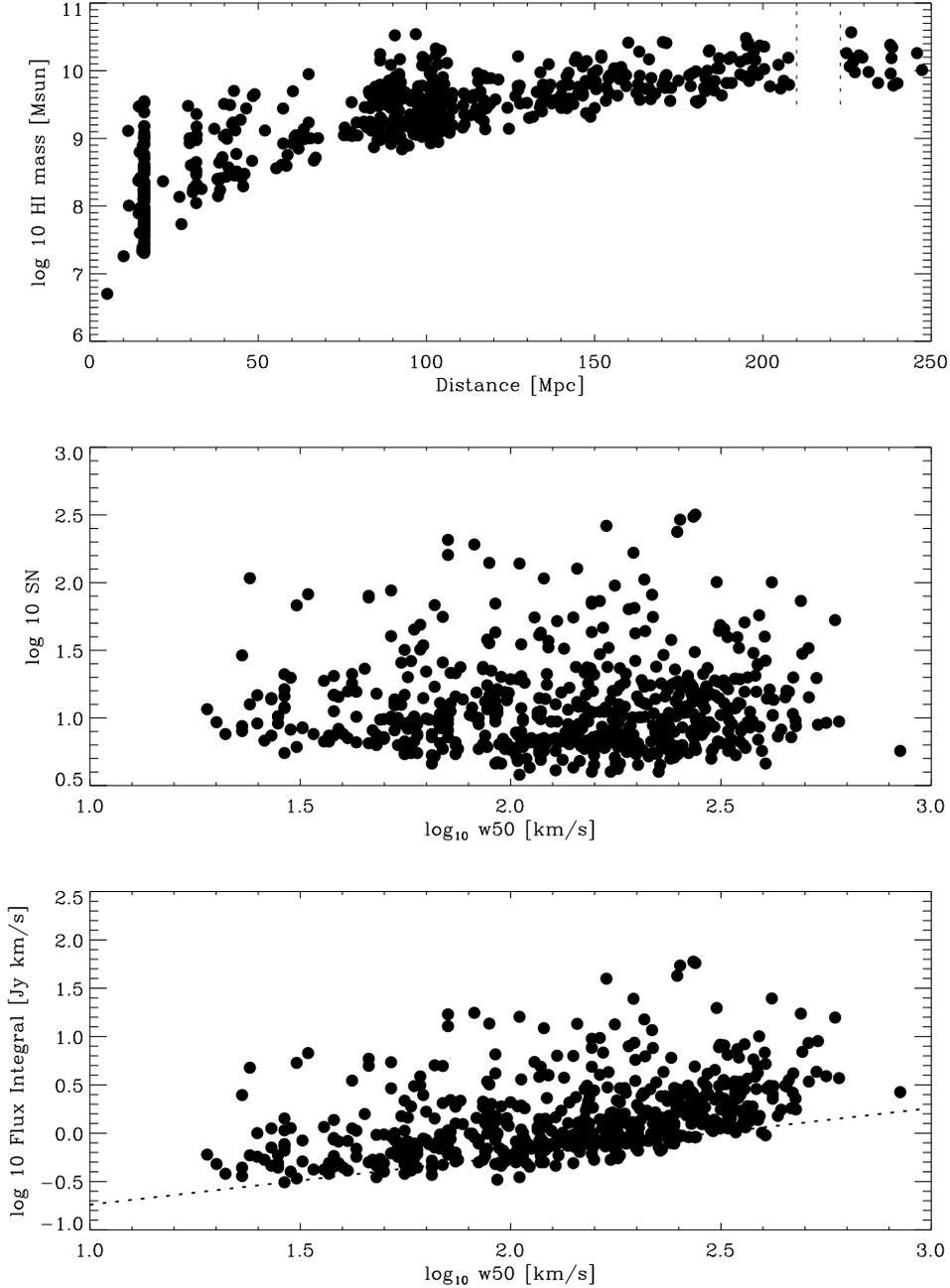}
\caption{Statistical properties of the Virgo south catalog.
The top panel shows the HI mass vs. distance for all sources
of type 1 and 2 presented in Table 1.  Note that galaxies
assigned to the Virgo cluster appear as the vertical
feature at 16.7 Mpc, and the gap at 220 Mpc is due
to RFI.  The middle panel shows the log$_{10}$ S/N ratio
versus velocity width W$_{50}$.  The lower envelope
is independent of S/N which indicates that
no significant bias is present in the detection
of sources of large width.  The bottom panel
shows the integrated flux versus velocity width W$_{50}$,
with the dashed line indicating a S/N limit of 6.5.}
\label{statistics}
\end{figure}

\begin{figure}[ht!]
\epsscale{1.0}
\plotone{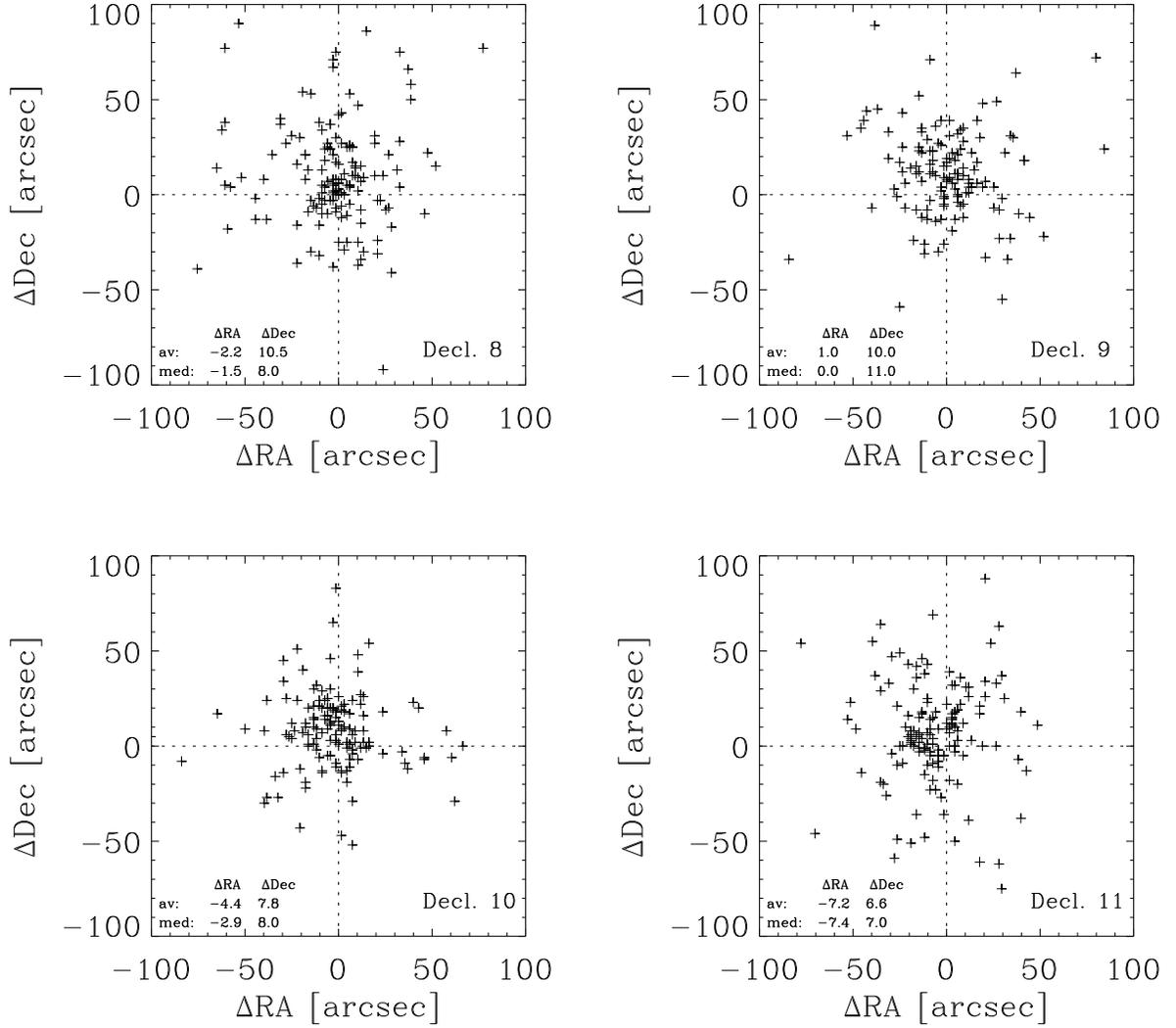}
\caption{Differences between the positions of the HI sources, as indicated by
the values embedded in the source names in Col. 1 of Table 1, and the
optical counterpart positions as listed in Col. 4 of Table 1. Sources
are separated by Declination bins of $1^\circ$: the label ``Dec 8'' identifies 
sources with Declination between +8$^\circ$ and +9$^\circ$, etc. Average 
and median offsets, expressed in arcseconds, are inset within each panel.}
\label{positions1}
\end{figure}

\begin{figure}[ht!]
\epsscale{1.0}
\plotone{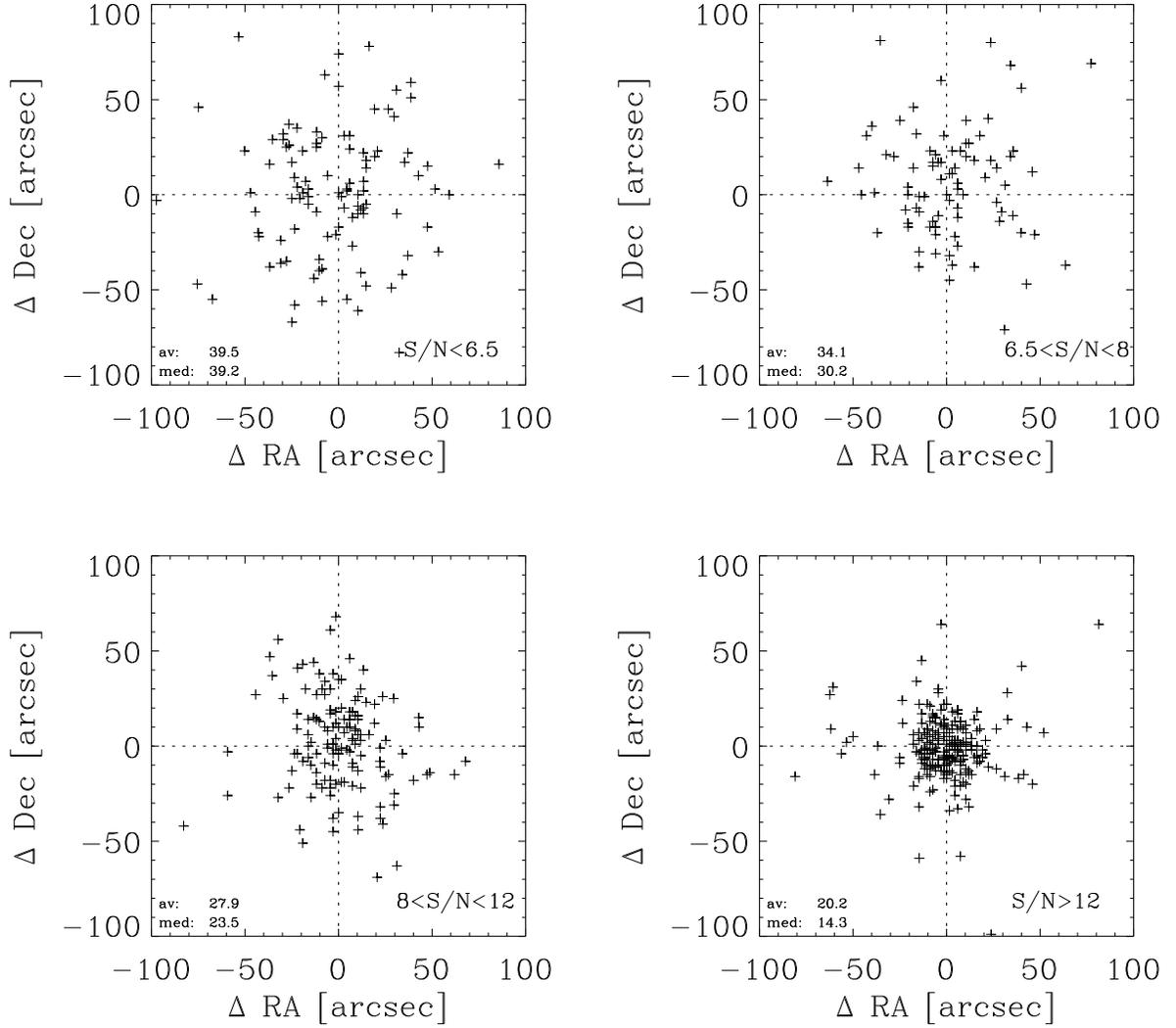}
\caption{Differences between the positions of the HI sources, as indicated by
the values in Col. 3 of Table 1, which are corrected for systematic telescope
pointing errors, and the
optical counterpart positions as listed in Col. 4 of Table 1. Sources
are separated by S/N as indicated in each panel. Average and
median offsets, expressed in arcseconds, are inset within each panel.
}
\label{positions2}
\end{figure}

\begin{figure}[ht!]
\plotone{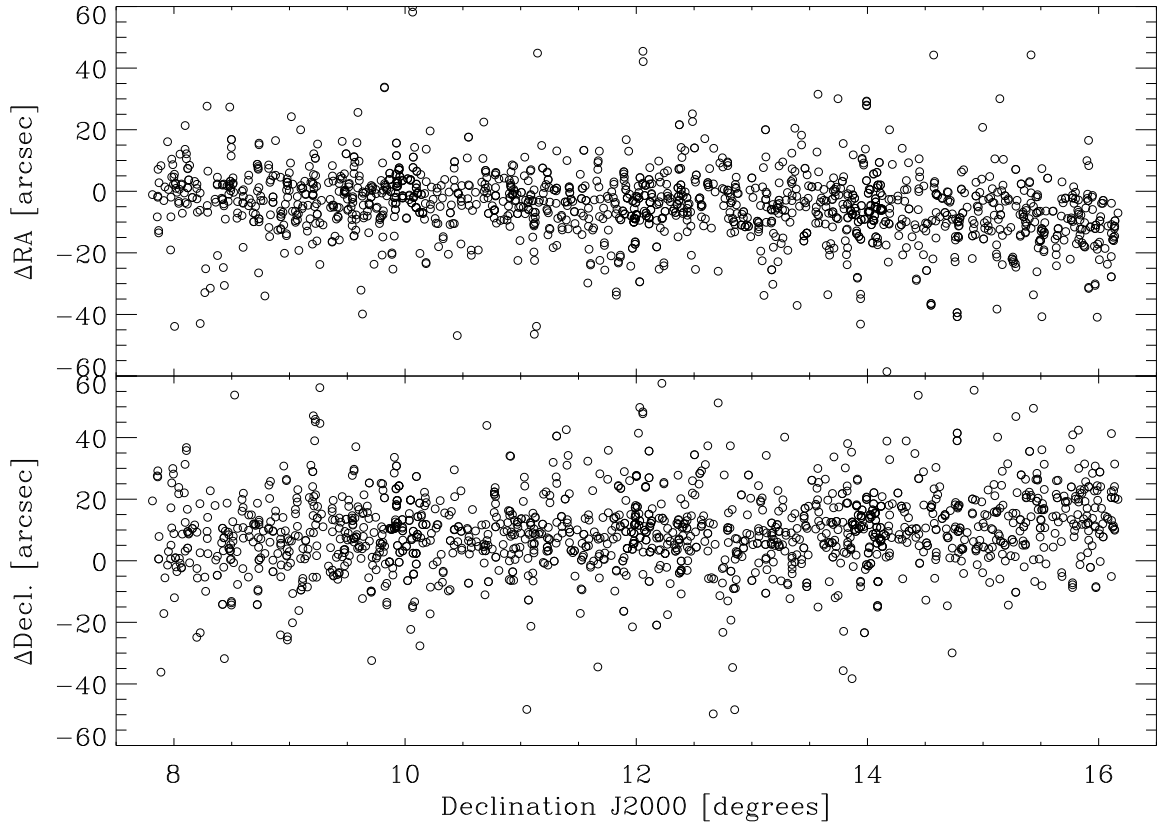}
\caption{Positional differences between 1568 ALFALFA and NVSS continuum
sources (Condon \etal~1998).  The offsets in the mean values of $\Delta$R.A. and $\Delta$Decl. 
are due to the Arecibo telescope pointing errors.}
\label{continuumall}
\end{figure}

\begin{figure}[ht!]
\plotone{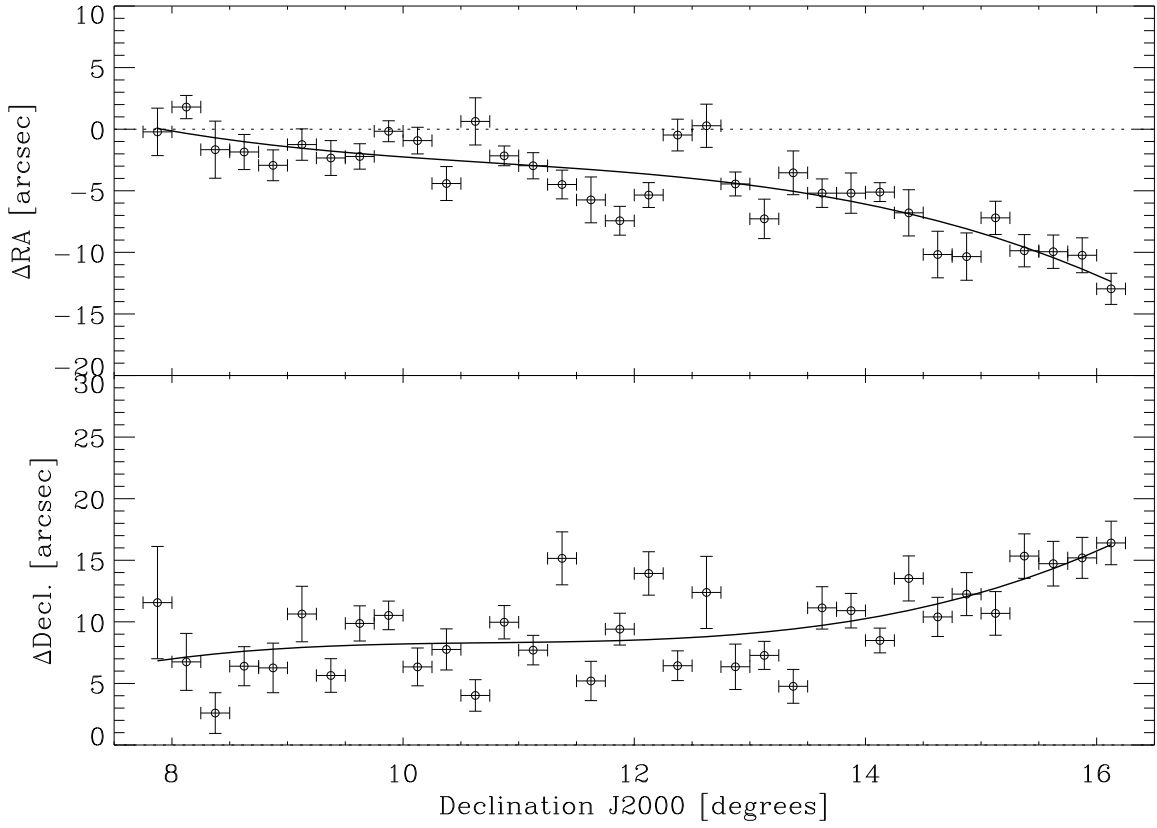}
\caption{Positional difference between
ALFALFA and NVSS continuum sources (Condon \etal~1998) averaged over 0.25 degree declination bins.
Vertical bars indicate the error on the mean for each bin.  Horizontal
bars indicate the bin width for a given point.
}
\label{continuumavg}
\end{figure}

\vfill
\end{document}